\def\ra{\rightarrow}
\def\decay{ $H_b\ra \tau\bar\nu X$ }
\def\bnu{{\bar\nu}}
\def\be{\begin{equation}}
\def\ee{\end{equation}}
\def\bea{\begin{eqnarray}}
\def\eea{\end{eqnarray}}

\def\nn{\nonumber\\}
\def\1{\cite{bksv}}
\def\r{\rho_{q}}
\def\w{\rho_{\tau}}
\documentstyle[prd,aps,preprint]{revtex}
\tighten
\begin{document}
\draft
\preprint{TPI-MINN-93/47-T; \ hep-ph/9311215}
\title{Nonperturbative Corrections to the Heavy Lepton Energy Distribution
in the Inclusive Decays $H_b\ra \tau\bar\nu X.$}
\author{L. Koyrakh}
\address{Theoretical Physics Institute,
University of Minnesota, Minneapolis, MN 55455}
\date{\today}
\maketitle
\begin{abstract}
Nonperturbative corrections up to $m_b^{-2}$ to the
heavy lepton energy distributions
are investigated in the  inclusive semileptonic weak decays of heavy
flavors in  QCD.
In the case of $B$-meson decays, for  $b \ra u\tau\bnu$  transitions
they decrease  the decay rate by 6\% of its perturbative value,
while for $b \ra c\tau\bnu$ they decrease it by 10\%.
\end{abstract}

\section{Introduction}
\label{intro}

In this paper we investigate the power corrections to the heavy lepton
energy distribution in the inclusive decays
$$H_b\ra \tau\bar\nu X,$$
where $H_b$ is a hadron containing heavy quark $b$
 and $\tau$ is the $\tau$-lepton.
We include nonperturbative corrections up to the order $1/m_b^2$, and
show that they cause the decay rate of $B$-mesons
to decrease by 6 to 10 percent of its perturbative value depending on the
mass of the quark in the final state.
The general approach of this work is based on the heavy quark mass
and the operator product expansions \cite{Isgu1}--\cite{Geor2}.
The massless lepton case was analyzed in the works
\cite{Bigi1}--\cite{Manohar}.
In those works the derivation of the corresponding matrix element and the
operator product expansion were discussed in details.
The lepton mass does not effect neither hadronic nor weak leptonic tensors.
Therefore we can use expressions (1)--(6)  and  (A1)--(A11) of
Ref. \cite{bksv}
for the hadronic invariant functions to get the corresponding matrix element.
What differs the decay with a heavy lepton from the decay with a massless
lepton is the phase space of the particles in the final state.
The present paper is a continuation of our previous work \cite{bksv}.
Here we will not repeat the discussion of applicability and interpretation
of the results. Everything said about it in \1 is valid in the cases
considered in this paper.

The first experimental observations of the decay were made
recently using the missing energy tag \cite{aleph}, \cite{beis}.
The missing energy
was associated with two $\nu_\tau$ in the decay chain
$b\ra\tau^-\bnu_\tau X, \  \tau^-\ra\nu_\tau X'$, which made it
difficult to reconstruct. The branching ratio was found to be
$4.08\pm0.76\pm0.62\%$ in \cite{aleph}  and $2.76\pm 0.47\pm 0.43 \%$
in \cite{beis} (more recent analysis), which is compatible
with the Standard Model.
However because of the difficulties in identification of
the decay mode \decay, the accuracy of the measurements
is still insufficient for direct comparison of the results
of this paper with the data.

\section{The Heavy Lepton Energy Distributions}
\label{dist}

As it was discussed in \1 in the semileptonic inclusive decays there are three
independent kinematical variables: $E_\tau$, $q_0$, and $q^2$,
where $E_\tau$ is the energy of the emitted lepton ($\tau$ in this paper),
$q_0$ is the energy and $q^2$ is the invariant mass of the lepton pair.
The fully differential distribution can be written as follows:
\be
{d\Gamma \over dE_\tau\,dq_0\,dq^2}=
        {1 \over 128\,\pi^2}\,|{\cal{M}}(E_\tau,q_0,dq^2)|^2.
\label{e:dgam}
\ee
The matrix element of the process ${\cal{M}}(E_\tau,q_0,dq^2)$
involves the same hadronic and leptonic tensors as used in \1 and which
are given by the expressions
(1)--(6)  and  (A1)--(A11) of Ref. \cite{bksv}. The corresponding phase
space and the boundaries for the kinematical variables for the massive
lepton case are briefly discussed in Appendix.

The double differential distribution in $E_\tau$ and $q^2$ could
be obtained from (\ref{e:dgam}) by integrating over $q_0$.
The corresponding expressions are somewhat cumbersome because
of the complicated limits of integration
over $q_0$ in $(q^2,q_0^2)$ plane.
Also there is little use of the double distribution because of
the difficulties in observing the considered decays at present time.
Therefore we will not write out the formulae for the double distributions
in $(q^2,E_\tau)$, but will rather integrate Eq. (\ref{e:dgam}) twice
and get the energy distribution for the charged lepton in the final state.
One more remark is in place here.
Although we are looking at a hadron decay, in our dynamical
approach we are considering a heavy quark
decaying in the external field created by its interactions with the
light degrees of freedom. That is why in the case at hand we have to use
the quark kinematical boundaries rather then the hadronic ones. This means
that in the formulas (\ref{e:gam})--(\ref{e:qmax}) we have to use $m_b$ and
$m_q$ instead of $M_{H_b}$ and $M_D$ correspondingly. We will use such a
kinematical boundary in the following calculations.
Our results then should be understood in the sense of
duality (see discussion in \1).

To get the energy distribution we integrate (\ref{e:dgam}) over $dq^2$ and
$dq_0$. In the dimensionless variables:
\be
 x={2\,E_\tau \over m_b}\,, \ \ \
        \r={m_q^2 \over m_b^2}\,, \ \ {\rm and} \ \ \w={m_\tau^2 \over m_b^2},
\ee
the result of the integration then takes the following form:
\begin{eqnarray}
\lefteqn{ \!\!\!\!\!\!{1 \over \Gamma_0}{d\Gamma \over dx}  =
  2\,\sqrt{x^2 - 4\,\w}\,
    \{ 3\,( 1 - \r + {\w} ) \,x - 2\,{x^2} - 4\,{\w} +
    \r\,( 3 - \r + 3\,\w ) \,f + }\nn
& &    ( -3 - \r + 6\,\w - \r\,\w - 3\,\w^2 ) \, f^2 +
        2\,(1 - \w ) ^2 \,f^3 + \nn
& & \!\!\!\!\!\! K_b\,[\,- {5\,x^2\over 3} + {14\,\w\over 3}  -
       {2\over 3}\,\r\,( 3 - \r + 3\,\w ) \,f +
        (1+{2\over 3}\,\r-4\,\w+{2\over 3}\,\r\,\w+\w^2)\,f^2 + \nn
& &     {2\over 3}\,(6 + \r - 6\,\w + 4\,\r\,\w - 6\,\w^2 +
             \r\,\w^2 + 6\,\w^3)\,{f^3\over \r} + \nn
& &   3\,(1-\w)^2\,(-1-2\,\r+2\,\w - 2\,\r\,\w -\w^2)\,{f^4 \over \r^2} +
        4\,(1 - \w )^4\,{f^5 \over \r^2}] + \nn
& & \!\!\!\!\!\! G_b\,[\,2\,x + {5\over 3}\,x^2 +
     4\,\r - {14\over 3}\,\w +
   (-2+3\,\r-{5\over 3}\,\r^2 - 2\,\w + 5\,\r\,\w)\,f +\nn
& &  ( -2 - {5\over 3}\,{\r^2} + 4\,{\w} + 8\,\r\,\w -
        {5\over 3}\,\r^2\,\w - 2\,\w^2 - 10\,\r\,\w^2)\,{f^2 \over \r} + \nn
& &    (-1 + \w ) \,
     (3 + {5\over 3}\,\r-8\,\w+{25\over 3}\,\r\,\w+5\,\w^2)\,{f^3\over \r} +
       5\,(1 - \w)^3\,{f^4\over \r}\,]\,\},
\label{e:taudist}
\end{eqnarray}
where
\be
        f={\r \over 1+\w-x},
\ee
and
\be
\Gamma_0=|V_{qb}|^2 \,{G_F^2 \,m_b^5 \over 192\,\pi^3}.
\ee
The quantities $K_b$ and $G_b$ are the following hadronic matrix elements:
\begin{eqnarray}
K_b & = &{1\over 2\,M_{H_b}}<H_b|{\bar b}\,(D^2-(v\cdot D)^2)\,b|H_b>/m_b^2 \nn
    & = &{1\over 2\,M_{H_b}}<H_b|{\bar b}\,{\vec\pi}^2 b |H_b>/m_b^2,
\label{eq:mupi}
\end{eqnarray}
\begin{equation}
G_b={1\over 2\,M_{H_b}}<H_b|\bar b \;{i \over2}\,
\sigma^{\mu\nu}G_{\mu\nu} b |H_b>/m_b^2,
\label{eq:muG}
\end{equation}
where $v$ is the 4-velocity of the decaying heavy meson $H_b$,
$D_\mu = \partial_\mu - i g A_\mu$, $A_\mu=A_\mu^a T^a$ is the gluon field
in matrix representation, $\vec\pi$ is the residual part of the heavy
quark momentum arising from its interaction with the
light degrees of freedom
and $G_{\mu\nu}=g G_{\mu\nu}^aT^a$ is the gluon strength tensor.

The energy distribution (\ref{e:taudist}) spans in $x$ from $x=2\,\sqrt{\w}$
to $x=1+\w-\r$. It includes the nonperturbative corrections -- terms
proportional to $K_b$ and $G_b$.
The part of eq.(\ref{e:taudist}) without
corrections coincides with the electron spectrum
in $\mu$ decay with a massive $\tau_\mu$ from ref. \cite{mu}.
In the limit $\r \ra 0$ we encounter the familiar
end-point singularities of the lepton spectrum.
To get the limit right we make the following substitutions:
\be
{f^n \over \r} \Rightarrow {\delta(1+\w-x) \over n-1}, \ \ \ n>1,
\label{e:rn-1}
\ee
and
\be
{f^n \over \r^2} \Rightarrow {\delta'(1+\w-x) \over (n-1)\,(n-2)}, \ \ \ n>2
\label{e:rn-2}
\ee
and only then take $\r$ to zero.
To see that this procedure gives the right limit, it is sufficient to
compare the results of integration of the both sides of (\ref{e:rn-1}) and
(\ref{e:rn-2}) in the limits $2\,\sqrt{\w}\leq x\leq 1+\w-\r$, multiplied by an
arbitrary integrable function and check that
the both sides are equal to each other.
The distribution for $\r \ra 0$ takes the form:
\begin{eqnarray}
\lefteqn{ \!\!\!\!\!\!{1\over\Gamma_0}{d\Gamma_{\r \rightarrow 0}\over dx} =
   \sqrt{x^2 - 4\,\w}\,\{\,( 6 + 6\,{\w} ) \,x - 4\,{x^2} - 8\,{\w}  + }\nn
& &  K_b\,[\,{28\,\w\over 3} +
     \delta'(1+\w-x)\,( -{1\over 3} +
        {{4\,{\w}}\over 3} - 2\,{\w^2} +
        {{4\,{\w^3}}\over 3} - {{{\w^4}}\over 3} )
       - {{10\,{x^2}}\over 3} ]  + \nn
& &  G_b\,[ - {28\,\w\over 3} +
     \delta(1+\w-x)\,( -{11\over 3} + 9\,{\w} -
        7\,{\w^2} + {5\,\w^3\over 3} )  +
     4\,x + {{10\,{x^2}}\over 3} ] \}.
\label{e:taudist0}
\end{eqnarray}

Now we can integrate distribution (\ref{e:taudist}) to get the
decay width including $1/m_b^2$ corrections:
\be
\Gamma  =
|V_{qb}|^2 \,{G_F^2 \,m_b^5 \over 192\,\pi^3} \{
( 1 +{G_b-K_b \over 2}\,)\,z_0(\r,\w) - 2\,G_b\,z_1(\r,\w)\,\},
\label{e:width}
\ee
where:
\bea
z_0(\r,\w) & = & \sqrt{\lambda}\,
  ( 1 - 7\,\r - 7\,{\r^2} + {\r^3} - 7\,{\w} - 7\,{\w^2} + {\w^3} +
 \r\,\w\,(12- 7\,\r - 7\,\w)\,)  + \nn
& &  12\,\r^2\,(1-\w^2)\,\log {(1 + v_q)\over (1-v_q) } +
  12\,\w^2\,(1-\r^2)\,\log {( 1 + v_\tau) \over (1-v_\tau)}
\eea
\bea
z_1(\r,\w) & =  & \sqrt{\lambda}\,
   (\,(1 - \r)^3 + (1-\w)^3 - 1 - \r\,\w\,(4 - 7\,\r - 7\,\w)\,) + \nn
& &    12\,\r^2\,\w^2\,\log{(1+v_q)(1+v_\tau) \over (1 - v_q) ( 1 - v_\tau)},
\eea
$$\lambda=\lambda(1,\r,\w)=1+\r^2+\w^2-2 \r -2 \w -2 \r \w,$$
$v_q$ and $v_\tau$ are the maximal velocities of the quark and the
$\tau$-lepton produced in the decay:
\be
v_q={\sqrt{\lambda} \over 1 + \r - \w},
 \ \ \ \
v_\tau={\sqrt{\lambda}\over 1 - \r + \w } .
\ee
There exists a simple relation between the two functions
$z_0(\r,\w)$ and $z_1(\r,\w)$ \cite{Bigi4} :
\be
z_1(\r,\w)=-2\,\r\,{dz_0(\r,\w) \over d\r}-2\,\w\,{dz_0(\r,\w) \over d\w} +
        4\,z_0(\r,\w)\,.
\ee
Width (\ref{e:width}) is symmetrical
function of $\r$ and $\w$ and
therefore its limit when $\r \ra 0$
\begin{equation}
\Gamma_{\r \rightarrow 0} =
|V_{qb}|^2 \,{G_F^2 \,m_b^5 \over 192\,\pi^3} \{
(1 +{G_b-K_b \over 2}\,)\,(1 - 8\,{\w} +  8\,{\w^3} - {\w^4} -
     12\,{\w^2}\,\log {\w}\,)  - 2\,G_b\,(1-\w)^4\,\}
\label{e:width0}
\end{equation}
is the same as limit when $\w \ra 0$ with
substitution $\r \Rightarrow \w$.

\section{Numerical estimates and experimental predictions}
\label{appl}

To make numerical estimates in this section we will consider the
$B$-meson decays $B \ra \tau\bnu X$.
We choose the following values for the parameters entering expressions
(\ref{e:taudist}), (\ref{e:taudist0}), (\ref{e:width}) and (\ref{e:width0})
(see discussion in \cite{Bigi4} and \1):
$m_b=4.8$ GeV, $m_c=1.4$ GeV, $m_u=0$, $m_\tau=1.78$ GeV,
$G_b=0.02$ and $K_b=0.017$ (note that for $\Lambda_b$ we have $G_b=0$
and the corrections are much smaller).

For the  width of the decay we see that the nonperturbative
corrections are negative. For $b\ra u\tau\bnu$ transitions they decrease
the width by 6\% of its perturbative value, while for the
$b\ra c\tau\bnu$ case they decrease
it by 10\% (note that for the massless lepton in the final state these
numbers are 4\% and 5\% correspondingly).
The quantity unsensitive to the uncertainties of known values of $V_{qb}$ is
$\Gamma(b\ra \tau\nu X)/\Gamma(b\ra e\nu X)$:
\be
r_q={\Gamma(b\ra \tau\nu X) \over \Gamma(b\ra e\nu X)}=
{z_0(\r,\w) \over z_0(\r,0)}\,[\,1-
2\,G_b\,({z_1(\r,\w) \over z_0(\r,\w)}-{z_1(\r,0) \over z_0(\r,0)})].
\label{e:r}
\ee
The first factor of last equation describes the phase space suppression
while the second contains the nonperturbative corrections.
The corrections reduce $r_c$ by 4\% and $r_u$
by 2\% of their perturbative values. Note that in the quantity $r_c$
the relative contribution of the corrections is
almost independent of the uncertainties
in the value of $c$-quark mass.

The obtained energy distributions (\ref{e:taudist}) and (\ref{e:taudist0})
could be applied to the decays involving
the transitions $b\ra c$ and $b \ra u$. The analysis of applicability
of the distributions was made in  \cite{Bigi4} and \1.
There it was shown that the proper quantity to confront
with experiment is
\be
\gamma(x)=\int^{1+\w-\r}_{x}dx'\,{1\over \Gamma_0}
        {d\Gamma(x') \over dx'},  \ \ \ \ 2 \sqrt{\w}\leq x<1+\w-\r.
\label{e:tilde}
\ee
This quantity does not contain the end-point singularities and is suitable
for direct comparison with the experimental data for $x$ not too close to
its maximal value (so that the operator product expansion is still valid
and we are not in the resonance region).
As we mentioned above, the hadronic kinematical region is different
from the quark one. Therefore to compare our results with experiment
the range of integration of the experimental distribution should
include the window between the the quark and hadronic boundaries
$1+\w-\r \leq x \leq M_B/m_b+\w-\r$. Correspondingly,
the reliable prediction can only be made
for $2\sqrt{\w}\le x \le x_{max}$, where $x_{max}=1+\w-\r - (M_B-m_b)/m_b$.
For $u$-quark $x_{max} \sim 1.05$, for $c$-quark $x_{max}\sim 0.95$.

The lepton energy spectrum is plotted on the Fig.1 for $b\ra c\tau\bnu$
and on the Fig.2 for $b\ra u\tau\bnu$. The delta-functions of
eq.(\ref{e:taudist0}) are not shown on the graph.
For comparison, on the same plots
we show the energy distributions for electrons in $b\ra ce\bnu$ and
$b\ra ue\bnu$ transitions correspondingly ($\w=0$).

The function $\gamma(x)$ is plotted on Fig.3 for the case
$b\ra c\tau\bnu$ and on Fig.4 for $b\ra u\tau\bnu$.
The solid line shows $\gamma(x)$
with the nonperturbative corrections while the dashed
line -- without them.

Although the functions are plotted for the whole range of $x$,
$2\sqrt{\w}\le x \le 1+\w-\r$, we can only trust the graphs for
$x < x_{max}$.

\section{Conclusions}
\label{conc}

We investigated the nonperturbative corrections
up to the order $1/m_b^2$ to semileptonic decays \decay and found
their contribution to the $\tau$-lepton energy spectrum and the
 width of the decay. In the case of $B$-meson decays the corrections
could be up to 10\% of the width.
Unfortunately at present time the experimental measurements
are not accurate enough in order to compare our results with the data.

\acknowledgments
The author would like to thank Prof. M. Shifman and Prof. A. Vainshtein
for many useful discussions.

\appendix
\section*{The Phase Space of the decay}
\label{pspace}

In this Appendix we briefly outline the derivation
of the Lorentz Invariant Phase Space (LIPS)
for the case of the inclusive semileptonic decay \decay.
Let us introduce the invariant kinematical variables in the following
way. Let $P$ be the 4-momentum of the decaying particle, so that
$M_{H_b}^2=P^2$;
$ q=p_\tau+p_{\bnu}$ is the 4-momentum of the lepton pair,
$p_\tau$ and $p_\bnu$ are  the momentums of the emitted charged
lepton and neutrino correspondingly.
We introduce $ m_X^2=p_X^2, $ -- the invariant mass squared of
the born hadronic state and $ m_{X\bnu}^2=p_{X\bnu}^2,$ -- the combined
mass of the hadrons and $\bnu$.
We have the following relations: $p_X=P-q$ and $p_{X\bnu}=P-p_\tau$.
The introduced invariant variables are related to the ones used
in \cite{bksv} in the following way:
\be
 m_X^2=M_{H_b}^2+q^2-2 \,M_{H_b}\,q_0,
\ee
\be
 m_{X\bnu}^2=M_{H_b}^2+p_\tau^2-2\,M_{H_b}\,E_\tau.
\label{e:m2e}
\ee

The inclusive decay rate in the normalization of \1
is given by the following expression:
\begin{equation}
\Gamma={1 \over 4}\,\int_{m^2_{X\,{min}}}^{m^2_{X\,{max}}} \, d m_X^2\,
        |{\cal M}(m_{X\bnu}^2,m_X^2,q^2)|^2\,
        d \Phi (P,p_X,p_{\tau},p_{\bnu}),
\end{equation}
where $|{\cal M}(m_{X\bnu}^2,m_X^2,q^2)|^2$ is
the matrix element which describes
the transition  into the hadronic states with mass $m_X^2$, it
is given by the expressions (2)--(8) in \cite{bksv}.  The lower
boundary of $m_X^2$ is given by the mass squared of the lightest final
hadronic state possible in the decay ($D$-meson mass for $b\ra c$ and
pion mass  for $b\ra u$ transitions ), while the upper
boundary of $m_X^2$ is $(M_{H_b}-m_\tau)^2$.
The $d \Phi (P,p_X,p_{\tau},p_{\bnu})$ is the three particle LIPS
for a particle with the 4-momentum $P$ going
into three particles with 4-momentums $p_X,p_{\tau}$ and $p_{\bnu}$.
It is given by the formulae:
\be
d \Phi (P,p_X,p_{\tau},p_{\bnu}) = (2\,\pi)^4\,
        \delta^4(P-p_X-p_{\tau}-p_{\bnu})\,\theta(p_X)\,\theta(p_\tau)\,
        \theta(p_{\bnu})\,{d^4p_X \over (2\,\pi)^3}\,
        {d^4p_{\bnu} \over (2\,\pi)^3}\,
        {d^4p_\tau \over (2\,\pi)^3},
\ee
where $\theta(a)$ denotes $\theta(a_0)$, and $a_0$ is the zero component
of the 4-vector $a$. For the three-particle phase space one can write:
\be
d \Phi (P,p_X,p_\tau,p_\bnu)=
        \int\,{d m_{X\bnu}^2\over 2\,\pi}\,d\Phi\,(P,p_{X\bnu},p_\tau)\,
        d\Phi\,(p_{X\bnu},p_X,p_\bnu)\,\delta(m_{X\bnu}^2-p_{X\bnu}^2)
\label{e:322}
\ee
where $d\Phi\,(a,b,c)$ is a two particle LIPS for a particle with
4-momentum $a$ going into particles with 4-momentums $b$ and $c$. The
equation (\ref{e:322}) gives the decomposition of a 3-particle LIPS
into 2-particle LIPSes.

In order to introduce a new variable $q^2$ into the phase space integral
one can multiply the equation (\ref{e:322}) by `1':
\be
\int \,dq^2\,\delta(q^2-(p_\bnu+p_\tau)^2)\,\theta(q_0)=1.
\label{e:1}
\ee
Performing the integrations over all variables except for
$m_X^2, q^2$ and $m_{X{\bnu}}^2$, which amounts to integrating the delta
functions in LIPSes along with determining the conditions for the delta
functions to be non-zeros, we arrive to the following formulae:
\be
\Gamma={1 \over 512\,\pi^2\,M_{H_b}^2}
        \int_{M_D^2}^{(M_{H_b}-m_\tau)^2}\,dm_{X{\bnu}}^2\,
        \int_{{M_D}^2}^{m_{X{\bnu}}^2}\,dm_X^2\,
        \int_{q^2_{min}(m_X^2,m_{X\bnu}^2)}^{q^2_{max}(m_X^2,m_{X\bnu}^2)}\,
        dq^2\,|{\cal M}(m_{X\bnu}^2,m_X^2,q^2)|^2,
\label{e:gam}
\ee
where
\be
q^2_{min}(m_X^2,m_{X{\bnu}}^2)={1 \over 2} [\,M_{H_b}^2+m_X^2-m_{X{\bnu}}^2+
        m_\tau^2-{M_{H_b}^2-m_\tau^2 \over m_{X{\bnu}}^2}\,m_X^2-
  {\sqrt{\lambda \,(M_{H_b}^2,m_\tau^2,m_{X{\bnu}}^2)}\,(m_{X{\bnu}}^2-m_X^2)
        \over  m_{X{\bnu}}^2} ],
\label{e:qmin}
\ee
\be
q^2_{max}(m_X^2,m_{X{\bnu}}^2)={1 \over 2} [\,M_{H_b}^2+m_X^2-m_{X{\bnu}}^2+
        m_\tau^2-{M_{H_b}^2-m_\tau^2 \over m_{X{\bnu}}^2}\,m_X^2+
  {\sqrt{\lambda(M_{H_b}^2,m_\tau^2,m_{X{\bnu}}^2)}\,(m_{X{\bnu}}^2-m_X^2)
        \over m_{X{\bnu}}^2} ],
\label{e:qmax}
\ee
and
\be
\lambda(a,b,c)=a^2+b^2+c^2-2\,a\,b-2\,a\,c-2\,b\,c.
\label{e:lamb}
\ee
The physical meaning of the function $\lambda(a,b,c)$ lies in
its relation to
the square of the spatial momentum $\vec p^{\,*\,2}$ of particles born in a
two - body decay in the center of  mass reference frame, for example:
\be
\vec p^{\,*\,2}_{X\bnu}=\vec p^{\,*\,2}_{\tau}=
{\lambda(M_{H_b}^2,m_\tau^2,m_{X{\bnu}}^2) \over 4\,M_{H_b}^2}.
\ee

The fully differential distribution (\ref{e:dgam}) in the invariant
variables looks as follows:
\be
{d\Gamma \over dm_X^2\,dm_{X\bnu}^2\,dq^2}=
        {1 \over 512\,\pi^2\,M_{H_b}^2}\,|{\cal{M}}(m_X^2,m_{X\bnu}^2,dq^2)|^2.
\label{e:digam}
\ee

%
\begin{figure}
\caption{The energy spectrum (\protect{\ref{e:taudist}}) of $\tau$
is plotted for $b\ra c\tau\bnu$ transitions.
The solid line shows the distribution
with the nonperturbative corrections, while the dashed line -- without them.
For comparison, on the same plot
we show the electron energy distribution for $b\ra ce\bnu$
transitions ($\w=0$).
The graph can only be trusted for $x<x_{max}\sim 0.95$.
}
\label{f:1}
\end{figure}
\begin{figure}
\caption{The energy spectrum (\protect{\ref{e:taudist0}}) of $\tau$
is plotted for $b\ra u\tau\bnu$ transitions ($\r=0$).
The solid line shows the distribution
with the nonperturbative corrections, while the dashed line -- without them.
For comparison, on the same plot
we show the electron energy distribution for $b\ra ue\bnu$
transitions ($\w=0$).
The graph can only be trusted for $x<x_{max}\sim 1.05$.
}
\label{f:2}
\end{figure}
\begin{figure}
\caption{The function $\gamma(x)$ plotted for the case $b\ra c\tau\bnu$.
The solid line shows $\gamma(x)$
with the nonperturbative corrections while the dashed line -- without them.
The graph can only be trusted for $x<x_{max}\sim 0.95$.
}
\label{f:3}
\end{figure}
\begin{figure}
\caption{The function $\gamma(x)$ for the case $b\ra u\tau\bnu$
($\r=0$). The solid line shows $\gamma(x)$
with the nonperturbative corrections while the dashed line -- without them.
The graph can only be trusted for $x<x_{max}\sim 1.05$.
}
\label{f:4}
\end{figure}

\begin{references}
%
\bibitem{aleph}
The ALEPH Collaboration, Phys. Lett. B {\bf 298}, 479 (1993)

\bibitem{beis}
A. Putzer, The ALEPH Collaboration, {\em Improved Measurement of the
$b\ra\tau\bnu_\tau X$ Branching ratio,}
Talk at the 5-th International Symposium on Heavy
Flavor Physics, Montreal, Canada, July 6-10, 1993

\bibitem{Isgu1}
N. Isgur, D. Scora, B. Grinstein and M. Wise,
Phys. Rev. D {\bf 39}, 799 (1989).

\bibitem{Wils1}
K. Wilson, Phys. Rev. {\bf 179}, 1499 (1969); \\
K. Wilson and J. Kogut, Phys. Rep. {\bf 12}, 75 (1974).

\bibitem{Volo1}
M. Voloshin and M. Shifman,  Yad. Fiz. {\bf 41}, 187 (1985)
[Sov. J. Nucl. Phys. {\bf 41}, 120 (1985)];
ZhETF {\bf 91}, 1180 (1986)
[{\em Sov. Phys. JETP} {\bf 64}, 698 (1986)].

\bibitem{Geor1}
E. Eichten and B. Hill, Phys. Lett. B {\bf 234}, 511 (1990);\\
H. Georgi, Phys. Lett. B {\bf 240}, 447 (1990).

\bibitem{Geor2}
J. Chay, H. Georgi and  B. Grinstein,  Phys. Lett. B
{\bf 247}, 399 (1990).

\bibitem{Bigi1}
I. Bigi, N. Uraltsev and A. Vainshtein,
Phys. Lett. B {\bf 293}, 430 (1992).

\bibitem{Bigi4}
I. Bigi, M. Shifman, N. Uraltsev and A. Vainshtein,
Phys. Rev. Lett. {\bf 71}, 496 (1993);
More detailed text is in preparation.

\bibitem{Bigi3}
I. Bigi, B. Blok, M. Shifman, N. Uraltsev and A. Vainshtein,
{\em ``A QCD `Manifesto' on Inclusive Decays of Beauty and Charm''},
Talk at DPF Meeting of APS, November 1992, Preprint TPI-MINN-92/67-T.

\bibitem{bksv}
B. Blok, L. Koyrakh, M. Shifman, A. Vainshtein,
Preprint NSF-ITP-93-68, TPI-MINN-93/33-T, hep/ph 9307247

\bibitem{Manohar}
A. Manohar and M. Wise, Preprint UCSD/PTH 93-14; CALT-68-1883;
hep-ph/9308246.

\bibitem{mu}
R.E.Shrock, Phys. Rev. D {\bf 24}, 1275 (1981)


\end{references}
\end{document}